# Coherent control of the silicon-vacancy spin in diamond

Benjamin Pingault[1], David-Dominik Jarausch[1], Christian Hepp[1], Lina Klintberg[1], Jonas N. Becker[2],

Matthew Markham[3], Christoph Becher[2], Mete Atatüre[1,*]

[1] Cavendish Laboratory, University of Cambridge, JJ Thomson Avenue, Cambridge CB3 0HE, UK
[2] Fachrichtung 7.2 (Experimentalphysik), Universität des Saarlandes, Campus E2.6, 66123 Saarbrücken, Germany
[3] Element Six Ltd., Global Innovation Centre, Fermi Avenue, Harwell Oxford, Didcot, OX11 0QR, UK

[*] Corresponding authors: bjpp2@cam.ac.uk and ma424@cam.ac.uk

**Abstract**

Spin impurities in diamond have emerged as a promising building block in a wide range of solid-state-based quantum technologies. The negatively charged silicon-vacancy centre combines the advantages of its high-quality photonic properties with a ground-state electronic spin, which can be read out optically. However, for this spin to be operational as a quantum bit, full quantum control is essential. Here, we report the measurement of optically detected magnetic resonance and the demonstration of coherent control of a single silicon-vacancy centre spin with a microwave field. Using Ramsey interferometry, we directly measure a spin coherence time, $T_2^*$, of 115 ± 9 ns at 3.6 K. The temperature dependence of coherence times indicates that dephasing and decay of the spin arise from single phonon-mediated excitation between orbital branches of the ground state. Our results enable the silicon-vacancy centre spin to become a controllable resource to establish spin-photon quantum interfaces.

**Introduction**

Spin-free materials, such as diamond and silicon, are ideal hosts to investigate and control the dynamics of spin-carrying defects. Among such systems, the nitrogen-vacancy centre in diamond has attracted great interest for electromagnetic field sensing [1,2,3,4], bio-labelling [5,6] and quantum information processing [7-11]. However, the nitrogen-vacancy centre only emits around 4% of its fluorescence into the zero-phonon line, limiting its use for quantum information processing in the absence of elaborate photonic structures such as photonic cavities [12]. In contrast, the negatively charged silicon-vacancy centre in diamond ($SiV^-$) emits around 80% of its photons into the zero-phonon line [13], with optical properties characterised by spectral stability and narrow inhomogeneous distribution in bulk diamond [14]. This makes the $SiV^-$ an ideal building block for a distributed quantum network [15, 16]. Furthermore, the $SiV^-$ offers two possible realisations of a quantum bit: One consists in using the two readily available orbital branches of the ground state, split by 50 GHz [17]. The other takes advantage of the $SiV^-$ ground state spin S = 1/2, optical signatures of which have previously been reported [18] and which promises longer coherence times, as estimated through coherent population trapping [19, 20]. Using the spin as a quantum bit allows to tune the resonance frequency with a magnetic field in the few GHz range, routinely used for NV centres. Achieving coherent control of this spin is a fundamental step for the implementation of spin-based quantum computing algorithms. Using microwave control provides more flexibility in the orientation of the magnetic field than all optical control which requires lambda-type transitions to be allowed. Furthermore, the capacity to address the spin state of the $SiV^-$ with microwave pulses is crucial to use the $SiV^-$ as an interface between optical and microwave photons for hybrid quantum computing [21].

Here, we report the first realisation of coherent control of the silicon-vacancy centre electronic spin. We perform optically detected magnetic resonance (ODMR) by tuning the frequency of a microwave field into resonance with the Zeeman splitting between two electronic spin states of a single $SiV^-$ centre. This allows to perform spectroscopy of the magnetic dipolar transitions between ground state levels of opposite electronic spin. By fixing the frequency of the microwave field in resonance with one of those transitions and varying the duration of the pulse, we coherently drive the $SiV^-$ electronic spin. We then perform Ramsey interferometry and measure a spin dephasing time of 115

± 9 ns at 3.6 K. The temperature dependence of the spin coherence times reveals that spin dephasing and population decay are dominated by single phonon excitation to the upper orbital branch of the ground state.

**Results**

**Spin initialisation and readout.** The SiV⁻ centre is composed of a silicon atom replacing two neighbouring carbon atoms in the diamond lattice. Its energy levels are characterised by orbitally split ground and excited states (Fig. 1a). We study single SiV⁻ centres in bulk diamond, created by implantation of isotopically purified $^{29}$Si followed by annealing (see the 'Methods' section). An external magnetic field lifts the degeneracy of the electronic spin S = ½, resulting in the energy level diagram shown in Fig. 1a. In our experiment, this magnetic field is applied at an angle of 109° with respect to the SiV⁻ symmetry axis, which dictates that all transitions between ground and excited states are optically allowed [19]. We use an optical pulse from a diode laser tuned to resonance with transition D1 (between ground state level 1 and excited state level D, red double arrow in Fig. 1a) to pump the SiV⁻ optically into the spin-down ground state. Fluorescence from the SiV⁻ is collected on the remaining transitions (solid grey arrows in Fig. 1a) following thermalisation among excited states (dashed grey arrows in Fig. 1a). The decrease of fluorescence during the laser pulse (Fig. 1b) allows to extract an initialisation fidelity of about 85% through a master equation model (see Supplementary). We perform the spin readout analogously by measuring the recovery of the leading edge of the fluorescence generated by a second optical pulse resonant with the same transition. The spin state measurement corresponds to the peak ratio between the leading edge of the readout pulse and that of the initialisation pulse. By varying the time delay between the two pulses, we extract a spin relaxation time $T_{1,\text{spin}}$ = 350 ± 11 ns (Fig. 1c) at 3.5 K, setting an upper time bound for subsequent driving of the SiV⁻ spin in this configuration.

**Optically detected magnetic resonance**. To address the electronic spin of the SiV$^-$, we apply a microwave pulse between the optical initialisation and readout pulses (Fig. 2a). Due to the nuclear spin ½ of the $^{29}$Si isotope, each electronic spin state displays a hyperfine splitting, as depicted in Fig. 2b. The microwave field drives the electronic spin, while preserving the nuclear spin state, which results in two distinct microwave resonance frequencies (orange and green circular arrows in Fig. 2b). When on resonance with one of the two transitions, the microwave transfers population from the initialised spin state to the depleted one, resulting in a peak in the ODMR spectrum (Fig. 2c). We measure the frequencies of the two resonances at different values of the applied magnetic field, as shown in Fig. 2d. We fit the observed frequency shift using an effective Hamiltonian described in Ref. [22], which includes the Jahn-Teller effect, spin-orbit coupling and electronic Zeeman effect, and accounts for the nuclear spin via the nuclear Zeeman effect and the hyperfine interaction (see Supplementary). From the fit, we extract a value of $A_\parallel = 70 \pm 2$ MHz for the hyperfine constant along the SiV$^-$ axis, in agreement with previously reported experimental [20, 23] and theoretical [24] values.

**Coherent control of the spin.** Having determined the frequencies of the electron spin-flipping transitions, we fix the microwave frequency in resonance with one of them and vary the duration of the microwave pulse, as illustrated in Fig. 3a. Figure 3b shows the evolution of the readout to initialisation peak ratio with respect to the microwave pulse duration when the microwave pulse is resonant with the transition between states $|\downarrow, n_\downarrow\rangle$ and $|\uparrow, n_\downarrow\rangle$ (the insert displays the same evolution for a microwave pulse resonant with the transition between states $|\downarrow, n_\uparrow\rangle$ and $|\uparrow, n_\uparrow\rangle$ ). This signal displays Rabi oscillations, sign of the coherent driving of the electronic spin of the SiV$^-$. The upward drift of the oscillations arises from spin-preserving population transfers between the two orbital branches of the ground state. This is confirmed by an eight-level master equation model (see Supplementary), used to reproduce the experimental curve (green and orange curves in Fig. 3b). This phenomenon highlights the importance of inter-branch dynamics at temperatures around a few kelvins. The detuned driving of the electron spin associated with the other nuclear spin orientation and of the transitions in the upper branch of the ground state (containing approximately one third of the population at 4 K) is responsible for a decreased contrast of the oscillations and their slightly irregular shape (See Supplementary). The linear dependence of the Rabi frequency with the square root of the microwave power (Fig. 3c), as well as its variation with respect to detuning (Fig. 3d) confirm the coherent nature of those oscillations. The observed bare Rabi frequency of about 15 MHz is comparable to what is obtained with NV$^-$ centres in similar conditions.

**Ramsey interferometry.** In order to measure the free induction decay time of the spin directly, we perform microwave Ramsey interferometry. We use a pulse sequence consisting of two π/2 microwave pulses separated by a variable delay (Fig. 4a). We fix the frequency of the microwave pulses such that in one case (Fig. 4b), the microwave frequency is in the middle of the two resonances, corresponding to a detuning of 27 MHz from each of them, and in the other case (Fig. 4c), the detuning from one resonance equals twice that from the other (36 MHz and 18 MHz, respectively). The measured peak ratio displays oscillations as a function of the free precession interval separating the two pulses. Since the frequency of Ramsey fringes corresponds to the value of the detuning of the microwave frequency, in Fig. 4b, the oscillations due to both resonances add constructively, while in Fig. 4c, they display a beating signal between two frequencies with one being twice the other. The decay of the Ramsey fringes gives a direct measurement of the dephasing time of the electronic spin $T_2^* = 115 \pm 9$ ns at 3.6 K. This value is commensurate with the characteristic time of population transfer between the two orbital branches of the ground state, taken as twice the orbital decay time $2T_{1,\text{orbital}} = 133 \pm 4$ ns at 3.6 K in our measurements.

**Single phonon-mediated spin dephasing and decay.** The strongest candidate for the cause of limited $T_2^*$ is expected to be phonon excitations from the lower to the upper orbital branches in the ground-state manifold. To verify this directly, we measure the spin dephasing rate $1/T_2$ as a function of temperature, as shown in Fig. 5a as blue dots. At the temperatures measured in the experiment, the orbital transition rate from the lower to the upper branch of the ground state is expected to be lower than that from the upper to the lower branch by a Boltzmann factor (see Supplementary). Since in our setup configuration, we underestimate the absolute temperature at the SiV, we take $1/2T_{1,orbital}$ as a reference for the orbital decay rate and display it as grey dots in Fig. 5a. The orbital decay rate increases linearly with temperature, a signature attributed to single-phonon excitation process between the two orbital branches of the ground state [20, 25]. The spin dephasing rate follows closely the orbital decay rate dependence on temperature, manifesting that spin dephasing is dominated by single-phonon-mediated transitions between the orbital branches. We also measure the temperature dependence of the spin decay rate $1/2T_{1,spin}$, as shown in Fig. 5b. Its linear dependence with temperature indicates that it is also a single phonon process. The suggested mechanism responsible for spin decay relies on excitation to the upper orbital branch of the ground state through the absorption of a single phonon. Due to a slight mismatch of spin quantisation axes between the two orbitals (see Supplementary), this excitation can result in a spin flip. This mechanism also explains why $T_{1,spin}$ can reach several milliseconds [20] when the magnetic field is aligned with the SiV symmetry axis which otherwise constitutes a competing quantisation axis for the spin via the spin-orbit interaction [19, 22]. The phonons responsible for those decoherence mechanisms have a frequency around 50 GHz corresponding to the orbital splitting of the SiV$^-$ and still have a significant population around 4 K.

**Discussion**

In conclusion, we have achieved coherent control of a single SiV$^-$ spin with a microwave field, following the identification of spin-flipping transitions through optically detected magnetic resonance. We have subsequently implemented Ramsey interferometry to get direct access to the spin dephasing time. The temperature dependence of the dephasing and decay rates reveal that both processes are governed by excitation from the lower to the upper orbital branch of the ground state through single phonon absorption. This indicates that the coherence times of the SiV$^-$ spin will improve significantly by cooling the system to lower temperatures. An alternative approach consists in splitting the orbital branches further apart by applying strain to the SiV⁻, thus increasing the energy required for phonons to cause decoherence. Those approaches would not necessarily improve significantly the coherence time of the orbital-based qubit, which is limited by phonon absorption and emission processes, simultaneously. The latter could be addressed by decreasing the phonon density of states around the frequency of the orbital splitting, which most likely requires nanostructures with dimensions smaller than about 120 nm, corresponding to half the phonon wavelength. Microwave control of the spin offers a strong advantage compared to all-optical control in that it gives much more flexibility for the orientation of the external magnetic field with respect to the SiV⁻ symmetry axis. In particular, control through microwave pulses allows to bring the magnetic field axis close to alignment with the SiV⁻ axis, thus giving access to cycling optical transitions [19] for single shot readout of the spin state [26]. The combination of the optical qualities of the SiV⁻ with the microwave addressing of some of its states also make the SiV⁻ a promising system to be used as a quantum transducer between optical and microwave photons [21]. Finally, the nuclear spin associated with $^{29}$Si can offer the opportunity of a quantum register, if it can be controlled coherently using radio frequency pulses and hyperfine coupling to the electron spin.

**Methods**

**Sample preparation.** The sample used for this experiment is a high pressure high temperature (HPHT) type IIa bulk diamond (Element Six) with surface oriented orthogonally to the [111] crystallographic axis. SiV⁻ centres were subsequently created by ion implantation of isotopically purified $^{29}$Si$^+$ ions at an implantation energy of 900 keV. Simulations using the stopping range of ions in matter algorithm [1] showed that the Si$^+$ stop at 500 ± 50 nm below the diamond surface. Implantation doses were tuned from one implantation site to the other from $10^9$ ions/cm$^2$ to $10^{12}$ ions/cm$^2$. Following the implantation, the samples were annealed at 1000 ◦C in vacuum for 3 hours, followed by an oxidation step in air for 1 hour at 460 ◦C. Single SiV$^-$ centres have been found in the 10$^{10}$ ions/cm$^2$ area.

**Experimental setup.** The diamond sample is mounted in a closed cycle liquid helium cryostat (Attodry 1000) reading 3.5 K at the sample and the temperature can be tuned via a resistive heater positioned under the sample mount or using heating caused by the microwave pulses. A superconducting coil around the sample space allows to apply a vertical magnetic field from 0 T to 9 T. The optical part of the setup consists of a home-built confocal microscope mounted on top of the cryostat and a microscope objective with NA=0.82 inside the sample chamber. The sample is moved with respect to the objective utilizing piezoelectric stages (Attocube) on top of which the sample is mounted. SiV$^-$ excitation is performed using a frequency tunable diode laser (Toptica DL pro design) whose frequency is maintained in resonance with an optical transition of the SiV$^-$ through continuous feedback from a wavemeter (High Finesse WSU). Optical pulses are generated with an acousto-optic modulator (AOM, AAOptoelectronic MT350-A0.12-800) controlled by a delay generator (Stanford Research Instruments DG645). The fluorescence from the SiV$^-$ is collected through the same objective. The laser light and SiV$^-$ fluorescence from the excited transition are filtered out using a home-built monochromator and the fluorescence from the remaining transitions is sent to an avalanche photodiode (PicoQuant Tau-Spad). The signal from the avalanche photodiode is sent to a time-to-digital converter (qutools quTAU) triggered by the delay generator. The microwave is generated by a frequency generator (Stanford Research Instruments SG384), pulses are generated with a switch connected to the delay generator used for optical pulses. Microwave pulses are then amplified by a microwave amplifier (Mini-Circuits ZHL-16W-43-S+). They travel through semi-rigid cables installed inside the cryostat. A single copper wire (20 μm diameter) positioned on top of the sample allows the microwave to be radiated to less than 20 μm away from the studied SiV$^-$.

**Acknowledgements**

We gratefully acknowledge financial support by the University of Cambridge, the ERC Grant PHOENICS, the Marie Skłodowska-Curie Actions Spin-NANO, Grant No. 676108, and the EPSRC National Quantum Technologies Programme NQIT EP/M013243/1. This research has been partially funded by the European Community's Seventh Framework Programme (FP7/2007-2013) under Grant agreement no. 611143 (DIADEMS). B. P. thanks Wolfson College (Cambridge) for support through a Research Fellowship. Ion implantation was performed at and supported by RUBION, the central unit of the Ruhr-Universität Bochum. We thank D. Rogalla for the implantation, C. Pauly for SIL fabrication. We thank M. Gündoğan, C. Stavrakas, A. Gali, J. Beitner, D. Kara, H.S. Knowles and Y. Kubo for helpful discussions.

**Author contributions**

M.A. and C.B. supervised the project. M.M., J.B. and C.B. prepared the sample. B.P., D.-D.J., C.H., L.K. and J.N.B. set up the experimental apparatus. B.P. performed the experiments and analysis, and designed the theoretical models.
All authors participated in the writing of the manuscript.

Correspondence and requests for materials should be addressed to B.P. and M.A. (bjpp2@cam.ac.uk and ma424@cam.ac.uk, respectively).

**Competing financial interests**

The authors declare no competing financial interests.

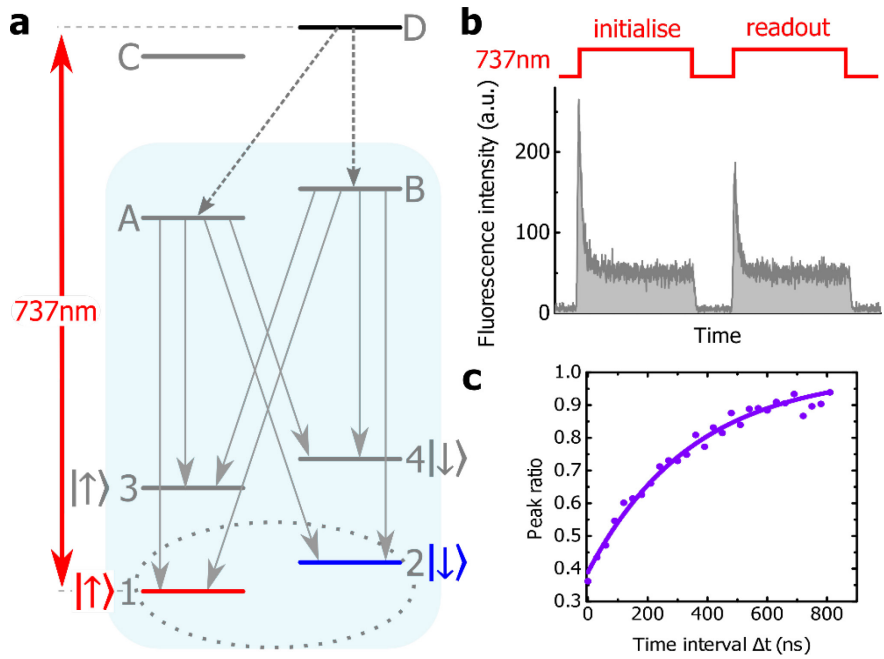

**Figure 1 | Spin initialisation and readout.** (**a**) Energy level scheme for a SiV⁻ under magnetic field. Excited state levels are labelled from A to D. States labelled 3 and 4 correspond to the upper orbital branch of the ground state. The dotted grey ellipse highlights the Zeeman-split lower orbital branch, on which we focus in this work and where state 1 (in red) corresponds to a spin up state and state 2 (in blue), to a spin down state. A resonant laser (red double arrow) drives transition D1 resonantly. Grey dashed arrows show fast decay paths from excited state D to excited states A and B. Solid grey arrows identify the measured fluorescent transitions. (**b**) A first optical pulse resonant with transition D1 causes optical pumping into the spin down state, as evidenced by the exponential decay of the fluorescence. The height of the leading edge of the fluorescence due to the second pulse indicates the recovery of the spin up population, and thus acts as a readout. (**c**) Recovery of the ratio between the leading edge of the readout pulse with respect to the initialisation pulse as a function of time at 3.5 K (purple dots). The purple curve is an exponential fit with 1/e value $T_{1,\text{spin}}$= 350 ± 11 ns.

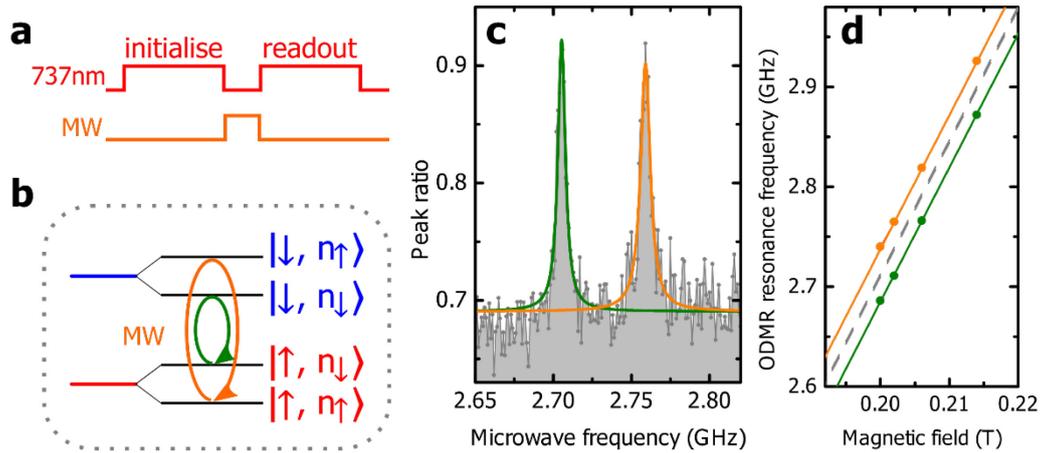

**Figure 2 | Optically detected magnetic resonance (ODMR).** (**a**) ODMR pulse scheme. A microwave pulse is applied in the interval between optical initialisation and readout pulses. (**b**) Close-up on the Zeeman-split lower branch of the ground state. The hyperfine interaction with the $^{29}$Si nuclear spin ½ causes a splitting of each electronic spin state. An applied microwave can flip the electronic spin, leaving the nuclear spin unchanged, resulting in two possible resonance frequencies (orange and green circular arrows). (**c**) ODMR spectrum: the peak ratio is plotted as a function of the microwave frequency. The two microwave resonances appear as two peaks, each fitted with a Lorentzian function [colours correspond to those of (**b**)] and separated by 54 MHz. (**d**) ODMR resonance frequencies as a function of external magnetic field (orange and green dots). The solid curves are fits based on the model developed in Ref. [22] expanded to include the nuclear spin. Orange and green curves correspond to the transitions in (**b**), the two overlapping dashed grey curves correspond to the transitions where both electronic and nuclear spins are flipped.

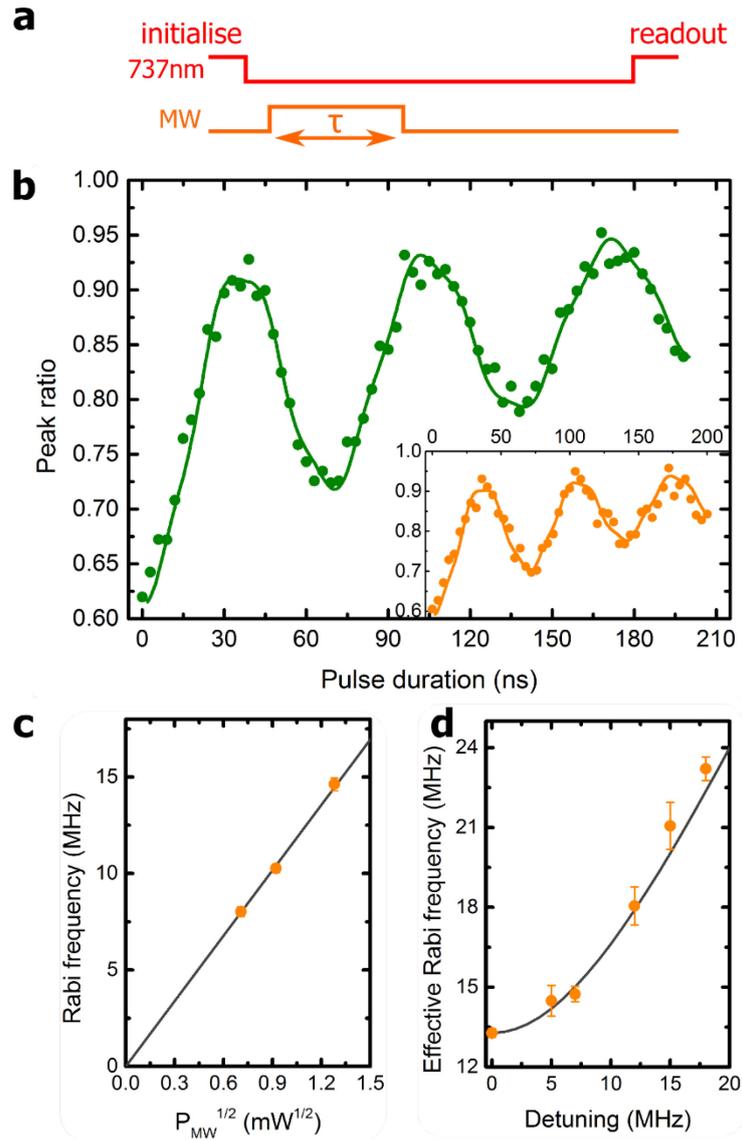

**Figure 3 | Rabi oscillations.** (**a**) Microwave and optical pulses scheme: the frequency of the microwave is fixed at one of the microwave resonances and its duration τ is varied. The time interval between initialisation and readout pulses is kept constant at 210 ns. (**b**) Rabi oscillations between states $|\downarrow, n_\downarrow\rangle$ and $|\uparrow, n_\downarrow\rangle$ [green circular arrow in Fig. 2b]: the measured peak ratio between leading edge of the readout and initialisation pulses oscillates as a function of microwave pulse duration (green dots). The green solid curve is a fit based on an eight-level master equation model (see Supplementary). The inset displays the measurement (orange dots) and fit (orange curve) of Rabi oscillations between states $|\downarrow, n_\uparrow\rangle$ and $|\uparrow, n_\uparrow\rangle$. (**c**) Evolution of the Rabi frequency as a function of the square root of the input microwave power $P_{MW}$, measured between states $|\downarrow, n_\uparrow\rangle$ and $|\uparrow, n_\uparrow\rangle$.

Error bars are smaller than dots. The grey line is a linear fit. (**d**) Effective Rabi frequency as a function of microwave frequency detuning, measured on transition $|\downarrow, n_\uparrow\rangle$ and $|\uparrow, n_\uparrow\rangle$ (orange dots). Error bars are the standard deviation of the Rabi frequencies extracted from fits. The grey line is a fit of the form $\sqrt{\Omega^2 + \delta^2}$, where $\Omega$ is the bare Rabi frequency and $\delta$ is the detuning.

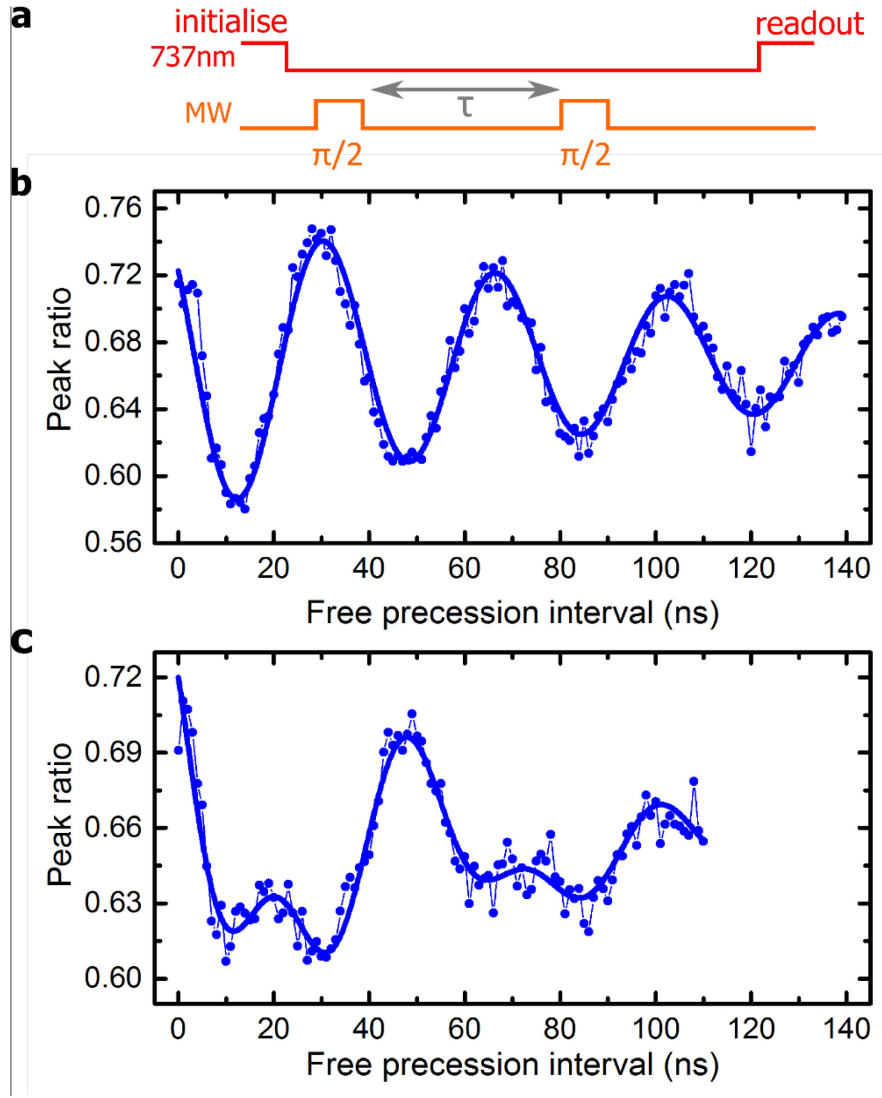

**Figure 4 | Ramsey interferometry.** (**a**) Microwave and optical pulses protocol: two microwave π/2 pulses are separated by a variable delay τ. The time interval between the initialisation and readout pulses is held constant. (**b**) Variation of the peak ratio as a function of the delay between the microwave pulses (blue dots) at 3.6 K. The microwave frequency is fixed at the average of the two resonance frequencies. The blue solid line is a fit of the form cos (2π·τ· δ) exp(-τ/$T_2^*$), where δ is the detuning between the microwave frequency and the two resonances. We extract $T_2^*$ = 115 ± 9 ns (The error corresponds to the standard deviation on the fit parameter). (**c**) The duration of the microwave pulses is kept unchanged compared to (**b**), but the frequency is tuned such that the detuning from one resonance is twice that with the other (36 MHz and 18 MHz, respectively). The

solid line is a fit where the cosine function in b is replaced by the sum of two cosine functions, with frequencies corresponding to the respective detunings.

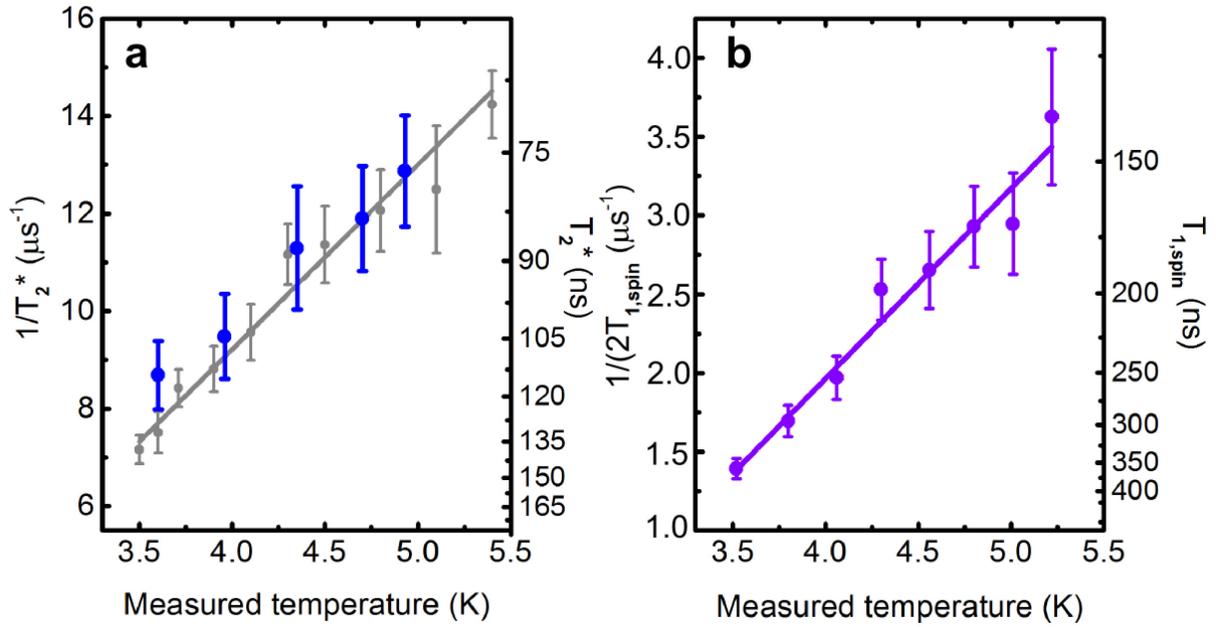

**Figure 5 | Phonon-induced spin dephasing and decay.** (**a**) Evolution of the spin dephasing rate $1/T_2^*$ (blue dots) and orbital population decay rate taken as $1/(2T_{1,\text{orbital}})$ (grey dots) as a function of temperature. The solid grey curve is a linear fit of the variation of $1/(2T_{1,\text{orbital}})$. (**b**) Temperature dependence of the spin decay rate $1/(2T_{1,\text{spin}})$. The solid purple curve is a linear fit of the experimental values. In (**a**) and (**b**), the error bars correspond to the standard errors on the fit parameter $T_2^*$, $T_{1,\text{orbital}}$ and $T_{1,\text{spin}}$, respectively. The temperature is measured below the sample mount, and, due to heating from the microwave, is expected to be lower than the temperature at the SiV$^-$.